\documentclass[5p,authoryear]{article}
\usepackage{xcolor}
\usepackage{graphicx}
\usepackage{amssymb,amsthm}
\usepackage[left=2.0cm,top=2.5cm,right=1.5cm,bottom=2.5cm]{geometry}

\begin{document}
%\begin{frontmatter}

%The role of synaptic transmission delays in spiking patterns and connecti{}vity of plastic neuronal networks
%Spiking patterns and its relation with the connectivity of a synaptic delayed plastic neuronal network
\section*{\bf
\noindent Plastic neural network with transmission delays\\ promotes equivalence between function and structure\\}

P. R. Protachevicz$^{1,2*}$, F. S. Borges$^{3,4}$, A. M. Batista$^{5,6}$, M. S. Baptista$^2$, I. L. Caldas$^1$, E. E. N. Macau$^7$, E. L. Lameu$^8$\\

{\small \noindent $^1$Physics Institute, University of S\~ao Paulo, S\~ao Paulo, SP,
Brazil.\\
$^2$Institute for Complex Systems and Mathematical Biology, SUPA, University of Aberdeen, Aberdeen, United Kingdom.\\
$^3$Department of Physiology and Pharmacology, State University of New York
Downstate Health Sciences University, Brooklyn, NY, USA\\
$^4$Center of Mathematics, Computation and Cognition, Federal University of ABC,
S\~ao Bernardo do Campo, SP, Brazil.\\
$^5$Post-Graduation in Science, State University of Ponta Gros\-sa, Ponta
Grossa, PR, Brazil.\\
$^6$Mathematics and Statistics Department, State University of Ponta Grossa,
Ponta Grossa, PR, Brazil.\\
$^7$Federal University of S\~ao Paulo, S\~ao Jos\'e dos Campos, SP, Brazil.\\
$^8$ Cell Biology and Anatomy Department, Hotchkiss Brain Institute, University
of Calgary, Calgary, Alberta, Canada.\\}
protachevicz@gmail.com\\

\noindent
The brain is formed by cortical regions that are associated with different cognitive functions. 
Neurons within the same region are more likely to connect than neurons in distinct regions, 
making the brain network to have characteristics of a network of subnetworks. The values of 
synaptic delays between neurons of different subnetworks are greater than those of the same subnetworks. 
This difference in communication time between neurons has consequences on the firing patterns 
observed in the brain, which is directly related to changes in neural connectivity, known as 
synaptic plasticity. In this work, we build a plastic network of Hodgkin-Huxley neurons in which 
the connectivity modifications follow a spike-time dependent rule. We define an internal-delay 
among neurons communicating within the same subnetwork, an external-delay for neurons belonging 
to distinct subnetworks, and study how these communicating delays affect the entire network dynamics. 
We observe that the neuronal network exhibits a specific connectivity configuration for 
each synchronised pattern. Our results show how synaptic delays and plasticity work together 
to promote the formation of structural coupling among the neuronal subnetworks. We conclude 
that plastic neuronal networks are able to promote equivalence between function and structure 
meaning that topology emerges from behaviour and behaviour emerges from topology, creating a 
complex dynamical process where topology adapts to conform with the plastic rules and firing 
patterns reflect the changes on the synaptic weights.     \\ 
Key-words: 
Synaptic plasticity; Hodgkin-Huxley model; Neural dynamics;
Synchronization

%%%%%%%%%%%%%%%%%%%%%%%%%%%%%%%%%%%%%%%%%%%%%%%
%%%%%%%%%%%%%%%%%%%%%%%%%%%%%%%%%%%%%%%%%%%%%%%
\section{Introduction}

Signal transmission delays are an intrinsic property of neuronal 
communication and are extremely relevant in brain activities 
\cite{petkoski2019,sreenivasan2019}. 
Asl et al. highlighted the importance of realistic time delays, 
emphasising that this property has 
just been marginally taking into account the used models over the
decades \cite{Asl2018}. 
The synaptic conduction delay is proportional to the axonal distance 
from the soma \cite{borges2022} and the experimental conduction 
velocity is 300 $\mu$m/ms in rat \cite{Stuart1997}. Therefore, 
neurons that are closer together have lower latencies, while 
communication between neurons in different hemispheres of the brain may 
have delays of a few hundred milliseconds \cite{Kosuke2022}. 
Lameu et al. \cite{lameu18} showed that small values of the
synaptic delay are related to synchronisation while non-trivial
topology is presented in networks with high delay.
However, in some cases, the synaptic delay can be associated with synchronisation 
suppression \cite{Mugnaine2018,Hansen2022}. Different 
transmission or synaptic delays can generate synchronous and 
asynchronous neuronal activities \cite{lubenov2008,protachevicz2020a}.

Maps of neuronal circuits obtained from functional magnetic resonance 
imaging (fMRI) show the existence of modules in brain networks 
\cite{Power2011}. Sporns and Betzel \cite{SpornsBetzel2015} discussed 
how this modular formation may be related to brain evolution favouring 
the emergence of functional specialisation and complex dynamics.
Lin et al. \cite{Lin2013} observed the fMRI and electrophysiological 
timing information delays during a visuomotor reaction-time task 
across five brain regions. Latency matrices of 9321 brain subregions 
from fMRI signals showed structures associated with different sensory 
states and brain functions \cite{Guo2022}.
Sun et al. demonstrated that certain intra and inter-networks 
delays can facilitate fast regular firings \cite{Sun2018}. 
Know and Choe \cite{Know2009} pointed to a possible role of 
facilitation dynamics to compensate such delays in motor 
neurons by a systematic approach.

Synaptic plasticity is the fundamental ability of the neurons to change their
connection intensities due to spike activities \cite{ramirez2016}. It has been
reported that memory and learning functions are supported by synaptic plasticity
in the brain \cite{abraham2019}.
Experimental data suggest that synchronous
spike dynamics patterns can be intensified due to the plasticity, being
associated with long-term memory processing. 
However, it is not completely
clear how plasticity and synchronisation are related \cite{fell2011}. 
Synaptic plasticity occurs by updating the connection strength depending 
on the difference of spiking times between pre and postsynaptic neurons. 
This function bounds topology with spiking time intervals and therefore 
with synchronisation. Different from networks that have fixed coupling, 
where synchronous behaviour emerges from topology, in the plastic network, 
topology emerges from behaviour and behaviour emerges from topology. 
This mutual influence creates a complex dynamical process where topology adapts to 
conform with the plastic rules and firing patterns reflect the changes 
on the synaptic weights that is not completely understood.  

Kim and Lim \cite{Kim2018a} studied spike synchronisation in the presence 
of excitatory spike time-dependent
plasticity (STDP). They observed a Matthew effect in synaptic plasticity due to a
positive feedback mechanism. In a similar framework, they also study the effect
of inhibitory STDP for the same network topology showing that both depression
and potentiation of inhibitory connections can occur \cite{Kim2018b}. Studying
the cerebellar ring network with synaptic plasticity, they still found that
phase, anti-phase, and complex out-of-phase activities are involved in the
long-term depression \cite{Kim2021}. 
Soltoggio and Stanley \cite{Soltoggio2012} reported  the relationship between local Hebbian
plasticity and learning using a computational approach focused on the noise and
weight saturation. 	

Aoki \cite{Aoki2015} demonstrated the self-organisation phenomena in a recurrent
network of oscillators in the presence of synaptic plasticity identifying phase,
anti-phase, coherent and chaotic activities. Phase, anti-phase and phase-lock
activities are the main types of synchronisation observed between brain regions 
\cite{tognoli2009,thatcher2012,carlos2020,Protachevicz2021}. 
Klimesch et al. \cite{klimesch2008} reported the importance of phase 
synchronisation in various cognitive processes. Phase synchronisation has been 
observed in distant cortical areas with long conduction delays \cite{knoublauch2003}. 
Some works also showed that time delay, synaptic types and connection 
densities play an important role in anti-phase synchronisation \cite{li2011,bodner1997}. 

In this work, we study how the emergence of synchronised symmetric patterns 
between the subnetworks depends on the absence and presence of long-term plasticity and on
internal and external transmission delays. 
The anatomy of neurons is directly related to the values of internal and 
external delays between subnetworks. The transmission velocity of the signal 
depends on the morphology of the dendrites and axons \cite{Manor1991, Boudkkazi2007} 
and the cell processing time \cite{Wang2005}. 
Moreover, the time required for neurons to communicate may be significantly extended
due to the physical distance between the sending and receiving cells \cite{Knoblauch2003}. 
Neurons in the same subnetwork have propagation delays associated with dendritic trees, 
with values ranging from submilliseconds to a few milliseconds \cite{AgmonSnir1993,Schierwagen2001}.
In connections between different subnetworks, the axonal propagation delay 
contributes greatly to signal latency \cite{Swadlow1987}. Considering that internal 
delays among neurons inside a subnetwork are smaller than the external ones between
subnetworks, we find that distinct patterns of synchronisation can be achieved
by changing the delays. In the presence of plasticity, one of main 
motivations is to investigate how the connections between the subnetworks 
evolve from initially random to a more structured configuration. In this context, 
we are able to identify the stronger connections between the subnetworks 
related to specific synchronised patterns. 

The increase of internal delays reduces the synchronous 
patterns inside the subnetworks. External delays, on the contrary, can promote 
collective synchronisation.  Subnetworks in one group symmetry whose neurons are 
connected with small external delays are less synchronous than 
those neurons connected via large delays. In addition to that, 
the synchronisation in one group is higher than for two, three, and four group symmetry. 
In particular, we note that different types of neuronal synchronisation are 
related to the network structure between the subnetworks.
Our results also suggest that synchronous behaviour reveals the structure 
being created due to the plasticity. More specifically, functional communities 
and their connections inferred by measurements of neural phase synchronisation 
reflect the subnetworks and their linkage structure provided by the structural 
topology of the synapses. 

The paper is organised as follows: In Section \ref{Section2}, we introduce our plastic 
neuronal network of coupled HH neurons and the diagnostic tool to identify synchronisation. 
In Section. \ref{Resuls}, we present the results of our study about the effects of synaptic 
delays between neuronal subnetworks. In the last section, we draw our conclusions.

\section{Plastic neuronal network}
\label{Section2}
\subsection{Hodgkin-Huxley model}

We consider the type-II neuron model proposed by 
\cite{hodgkin52}. The individual dynamics of each Hodgkin-Huxley (HH) neuron in
the network is given by 
\begin{eqnarray}
	C\dot{V_i} & = & I_i-g_{\rm K}n_i^4(V_i-E_{\rm K})\nonumber \\
	& & -g_{\rm Na}m_i^3h_i(V_i-E_{\rm Na})-g_L(V_i-E_{\rm L}) \nonumber \\
	& & +(V_{\rm r}^{\rm +}-V_i)\sum_{j=1}^{N}g_{ij} f_j(t-\tau_{ij}), \\
	\dot{n}_i & = & \alpha_{n_i}(v_i)(1-n_i)-\beta_{n_i}(v_i)n_i,\\
	\dot{m}_i & = & \alpha_{m_i}(v_i)(1-m_i)-\beta_{m_i}(v_i)m_i,\\
	\dot{h}_i & = & \alpha_{h_i}(v_i)(1-h_i)-\beta_{h_i}(v_i)h_i,\\
	\dot{f}_i & = & \frac{-f_i}{\tau_{\rm s}}. 
\end{eqnarray}
Equation (1) represents the membrane dynamics of the neuron $i$. $C$
($\mu$F/cm$^2$) is the membrane capacitance and $I_i$ ($\mu$A/cm$^2$) is a
constant current density chosen in the interval $[10,11]$. Depending on the $I_i$ value, 
the HH model can exhibit silence, bistability, and repetitive spike firings 
\cite{Luccioli2006,Pospischil2008,Giannari2020,Shi2016}. Once one understands 
well how bifurcations and dynamical behaviour happens as a parameter is changed 
in a nonlinear dynamical system, it is natural to take that parameter as the one 
to induce heterogeneity in a network. Following this idea, we have considered the 
heterogeneity in the constant current in the regime of repetitive spike firings 
as done in the references \cite{Popovych2013,Borges2015,Borges2016,Lameu2018}.
The parameters $g_{\rm K}$, $g_{\rm Na}$, and $g_{\rm L}$ are the conductance 
of the potassium, sodium, and leak ion channels, respectively. $E_{\rm K}, E_{\rm Na}$, and
$E_{\rm L}$ are the reversal potentials for these ion channels. $V_{\rm r}^{\rm +}$
corresponds to the excitatory reversal potential. $g_{ij}$ is the excitatory
coupling strength from the presynaptic neuron $j$ to the postsynaptic neuron
$i$ with maximum and minimum value within the interval $[0.0,0.01]$. We consider
that the neuron has no self-connections, implying $g_{ii}=0$. $f_i(t)$ is
the normalised synaptic current from the neuron $j$ to $i$. The state variable
$f_i$ decays exponentially and it is updated to the unity ($f_i\to 1$)
at the spike time $t_i$ of the neuron $i$ \cite{Rothman2014,borgesNN2017}. Equation (5)
corresponds to an exponential decay on the evolution of $f_i(t)$. The parameter
$\tau_{\rm s}$ is the synaptic time decay and $\tau_{ij}$ the delay on the
synaptic transmission \cite{Asl2017,Markram1997,Markram1997b}. The intensity of
synaptic current with time delay on the signal transmission depends
on the state of the pre and postsynaptic neuron. $\tau_{ij}$ assumes values
$\tau_{\rm int}$ and $\tau_{\rm ext}$ for internal and external connections between
the subnetworks, respectively. In Eqs. (2) and (3), the functions $m(v_i)$ and
$n(v_i)$ represent the sodium and potassium activation, respectively. In
Eq. (4), $h(v_i)$ is the function for sodium inactivation. The functions
$\alpha_n$, $\beta_n$, $\alpha_m$, $\beta_m$, $\alpha_h$, and $\beta_n$ are
given by
\begin{eqnarray}
	\alpha_n(v) & = & \frac{0.01v+0.55}{1-\exp\left(-0.1v-5.5\right)},\\
	\beta_n(v) & = & 0.125\exp\left(\frac{-v-65}{80}\right),\\
	\alpha_m(v) & = & \frac{0.1 v+4}{1-\exp\left(-0.1v-4\right)},\\
	\beta_m(v) & = & 4\exp\left(\frac{-v-65}{18}\right),\\
	\alpha_h(v) & = & 0.07\exp\left(\frac{-v-65}{20}\right),\\
	\beta_h(v) & = & \frac{1}{1+\exp\left(-0.1v-3.5\right)},
\end{eqnarray}
where $v=V/[{\rm mV}]$. In our simulations, we consider $C=1$ $\mu$F/cm$^{2}$,
$E_{\rm Na}=50$ mV, $E_{\rm K}=-77$ mV, $E_{\rm L}=-54.4$ mV, $g_{\rm Na}=120$
mS/cm$^2$, $g_{\rm K}=36$ mS/cm$^2$, $g_{\rm L}=0.3$ mS/cm$^2$, and
$\tau_{\rm s}=2.728$ ms. The reversal potential for excitatory connections is 
$V^{\rm +}_{\rm r}=20$ mV \cite{borges17}. For the numerical integration, we
use the Runge-Kutta fourth-order method with a time step equal to
$\delta t=0.01$ ms. 

%%%%%%%%%%%%%%%%%%%%%%%%%%%%%%%%%%%%%%%%%%%%%%%

\subsection{Spike-time dependent plasticity}

Spike-time dependent plasticity (STDP) is a process that produces changes in the
synaptic strength. It is calculated taking into consideration the times between
the spikes of the postsynaptic neuron $t_i$ and the presynaptic neuron $t_j$. 
For  each synaptic connection, the presynaptic neuron is that one which sends 
the signal while the postsynaptic neuron receives such signal. The intensity 
of the excitatory synaptic weight is defined by the coupling strength.
The change in the excitatory synaptic weights $\Delta g_{ij}$ due to the time
difference $\Delta t_{ij}=t_i-t_j$ is given by
\cite{bi98,Markram2012,Caporale,Popovych}
\begin{equation}
	\Delta g_{ij}= \left\{
	\begin{array}{ll}
		\displaystyle A_1{\rm e}^{(-\Delta t_{ij}/\tau_1)}\;\;\;,\;\;\; {\rm if} \;\;\;
		\Delta t_{ij} \ge 0\;\;\;\;\\
		\displaystyle -A_2{\rm e}^{({\Delta t_{ij}/\tau_2})}\;\;,\;\;\; {\rm if} \;\;\;
		\Delta t_{ij}<0 \\
	\end{array} \label{STDP} \right .
\end{equation}
where $A_1=1$, $A_2=0.5$, $\tau_1=1.8$ ms, and $\tau_2=6$ ms. The synaptic
weights are updated according to Eq. (\ref{STDP}), where 
$g_{ij}\rightarrow g_{ij}+G\cdot\Delta g_{ij}$. The change rate of the synaptic
weight is considered as $G=10^{-5}$ mS/cm$^{2}$. The initial value of all
excitatory synaptic weights is given by $g_{ij}=0.001$ mS/cm$^2$. The minimal 
and maximal excitatory synaptic weight are considered in the 
interval [$g_{\rm min}$, $g_{\rm max}$] = [0,0.1] nS.

\begin{figure}[ht!]
	\begin{center}
		\includegraphics[scale=0.1]{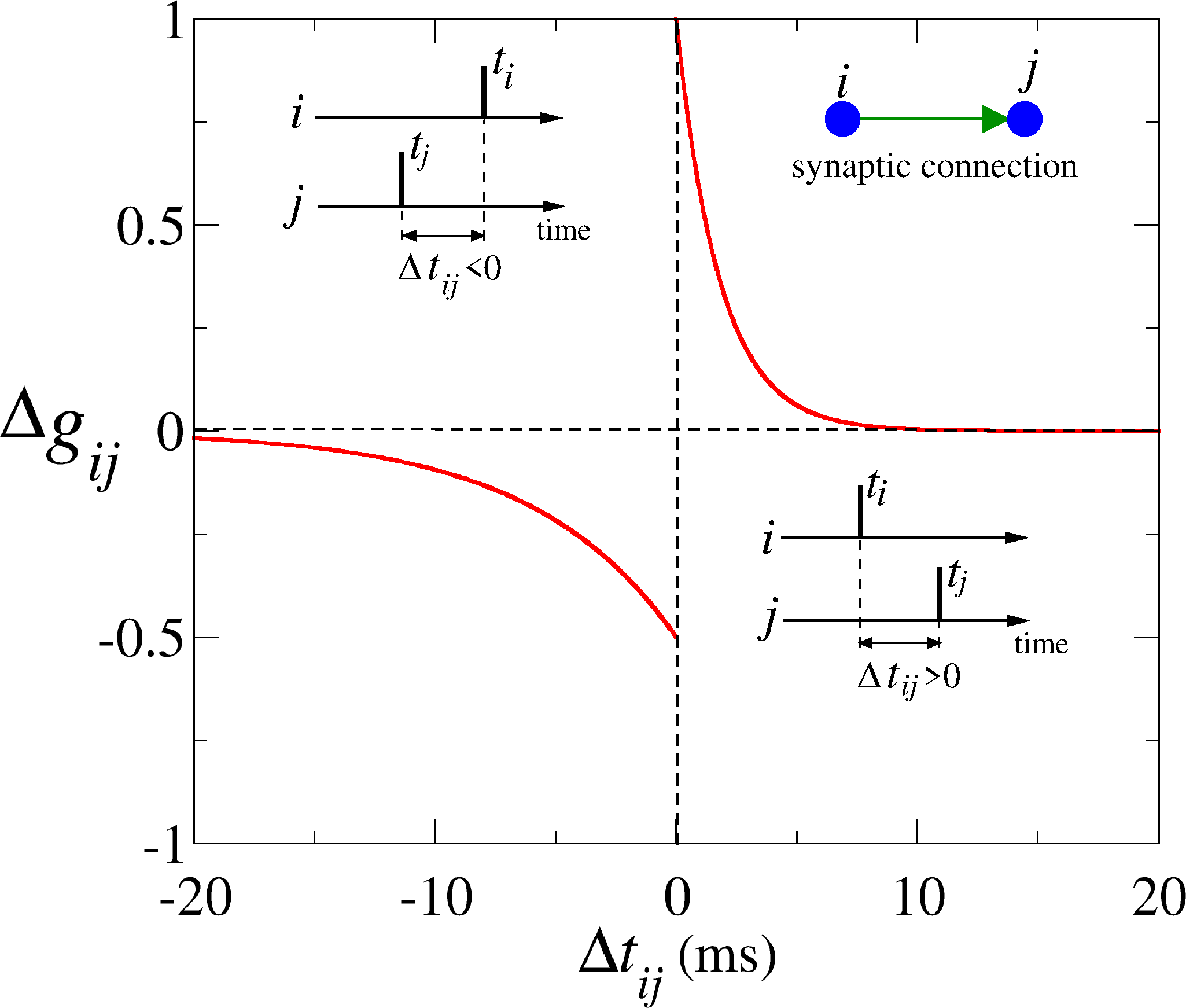}
		\caption{Plasticity curves as a function of $\Delta t_{ij}$ for excitatory 
			synapses. Negative and positive plus zero values of $\Delta t_{ij}$ correspond 
			to synaptic depression and potentiation, respectively. The inset shows a 
			schematic representation of the calculation of time difference $\Delta t_{ij}$ 
			for a connection from neuron $i$ to neuron $j$, for the case of synaptic 
			depression ($\Delta t_{ij}<0$) and synaptic potentiation ($\Delta t_{ij}\ge0$).}
		\label{fig1}
	\end{center}
\end{figure}

Figure \ref{fig1} displays the plasticity curves described by Eq. ({\ref{STDP}})
(red line) as a function of $\Delta t_{ij}$. This figure shows that the difference 
on the spike times in which plastic rule generates synaptic depression or synaptic 
potentiation. High values of the time difference generate synaptic changes 
close to zero, as shown in the figure. To exemplify such update protocol, 
suppose that there is a synaptic connection from the neuron $i$ to neuron $j$, 
as shown in the inset of Figure \ref{fig1}. Neurons $i$ and $j$ are represented 
by the circles while the synaptic connection from neuron $i$ to neuron $j$ is 
represented by the arrow.	 If neuron $j$ spikes before the neuron $i$, 
$\Delta t_{ij}$ will be negative, and the change in the synaptic connection 
from neuron $i$ to neuron $j$ ($\Delta g_{ij}$)  will be negative 
(for small $|\Delta t_{ij}|$) or zero (for large $|\Delta t_{ij}|$). 
In other way, considering the same synaptic connection from neuron $i$ 
to neuron $j$, but with neuron $i$ spiking before that neuron $j$, $\Delta t_{ij}$ 
will be positive, and the change in the synaptic connection ($\Delta g_{ij}$) 
will be positive (for small $|\Delta t_{ij}|$) or zero (for large $|\Delta t_{ij}|$).  

%%%%%%%%%%%%%%%%%%%%%%%%%%%%%%

\subsection{A network of subnetworks}

We consider $N=400$ non-identical HH neurons separated into $S=4$ subnetworks
with $N_{\rm sub}=100$ neurons each one ($N=S\cdot N_{\rm sub}$). The heterogeneity
in the system is given by the neuron currents $I_i$. To facilitate the
visualisation and interpretation of our results, we sorted neurons in each
subnetwork in ascending order according to their spiking frequency (or $I$).
Therefore, the neuron $i=1$ has the slowest spiking frequency and the neuron
$i=100$ has the highest one. The neurons are connected by means of excitatory
synapses. For the initial coupling configuration, each subnetwork has an
internal all-to-all topology without self-connections (autapses) \cite{Protachevicz2020b}. 
The connections between subnetworks or external ones are randomly distributed with 
a certain probability. Thus, the internal and external probability of connections
are given by $p_{\rm int}=1$ and $p_{\rm ext}=0.05$, respectively
\cite{Schmidt,Johnson}. New connections are not allowed between subnetworks,
however, changes in the weights of initial external connections are permitted. 
The network does not evolve to a configuration of only one community due to the
plasticity due to the fixed internal connection probability 
between the subnetwork. With regard to the subnetworks, we consider an internal and external
transmission delay given by $\tau_{\rm int}$ and $\tau_{\rm ext}$, respectively. 

%%%%%%%%%%%%%%%%%%%%%%%%%%%%%%%%%%%%%%%%%%%%%%%

\subsection{Measuring synchronisation and symmetries}

In order to study neuronal synchronisation and symmetries, we compute the order
parameter. Firstly, we use the traditional Kuramoto order parameter as a
diagnostic tool for the whole network, that is given by
\cite{kuramoto84} 
\begin{equation}
	R_{\rm T}(t)=\left|\frac{1}{N}\sum_{j=1}^{N}e^{{\rm i}\phi_j(t)}\right|,
	\label{Rtempo}
\end{equation} 
where ``${\rm i}$'' is the imaginary unit $\sqrt{-1}$ and $\phi_j(t)$ is the
neural phase associated with the spikes of each neuron $j$, given by
\begin{equation}
	\phi_j(t)=2\pi\frac{t-t_{j,k}}{t_{j,k+1}-t_{j,k}},
\end{equation}
$t_{j,k}$ is the time when a $k$-th spike ($k=0,1,2,\dots$) happens in the neuron
$j$ ($t_{j,k}<t<t_{j,k+1}$). 

The time-average order parameter for the network is given by
\begin{equation}
	{\bar{R}}=\frac{1}{t_{\rm fin}-{t_{\rm ini}}}\sum_{t_{\rm ini}}^{t_{\rm fin}}
	R_{\rm T}(t), \label{Rmedio}
\end{equation} 
in which $t_{\rm fin}-t_{\rm ini}$ is the time window set to measure the phases,
where $t_{\rm ini}$ and $t_{\rm fin}$ correspond to the initial and final time of
the analyses, respectively. In our simulations, we consider $t_{\rm ini}=80$ s and
$t_{\rm fin}=100$ s. The magnitude of the time-average order parameter tends to
the unity when the network has a globally synchronised behaviour. For
uncorrelated spiking phases, the order parameter is close to $0$.

The traditional Kuramoto order parameter for each subnetwork, $s=1,2,3,$ and
$4$, is described as
\begin{equation}
	R_{(s)}(t)=\left|\frac{1}{N_{\rm sub}}\sum_{j=(s-1)\cdot N_{\rm sub}+1}^{s\cdot N_{\rm sub}}
	e^{{\rm i}\phi_j(t)}\right|. \label{Rsubnetwokrtempo}
\end{equation} 

\begin{figure*}[ht!]
	\begin{center}
		\includegraphics[scale=0.36]{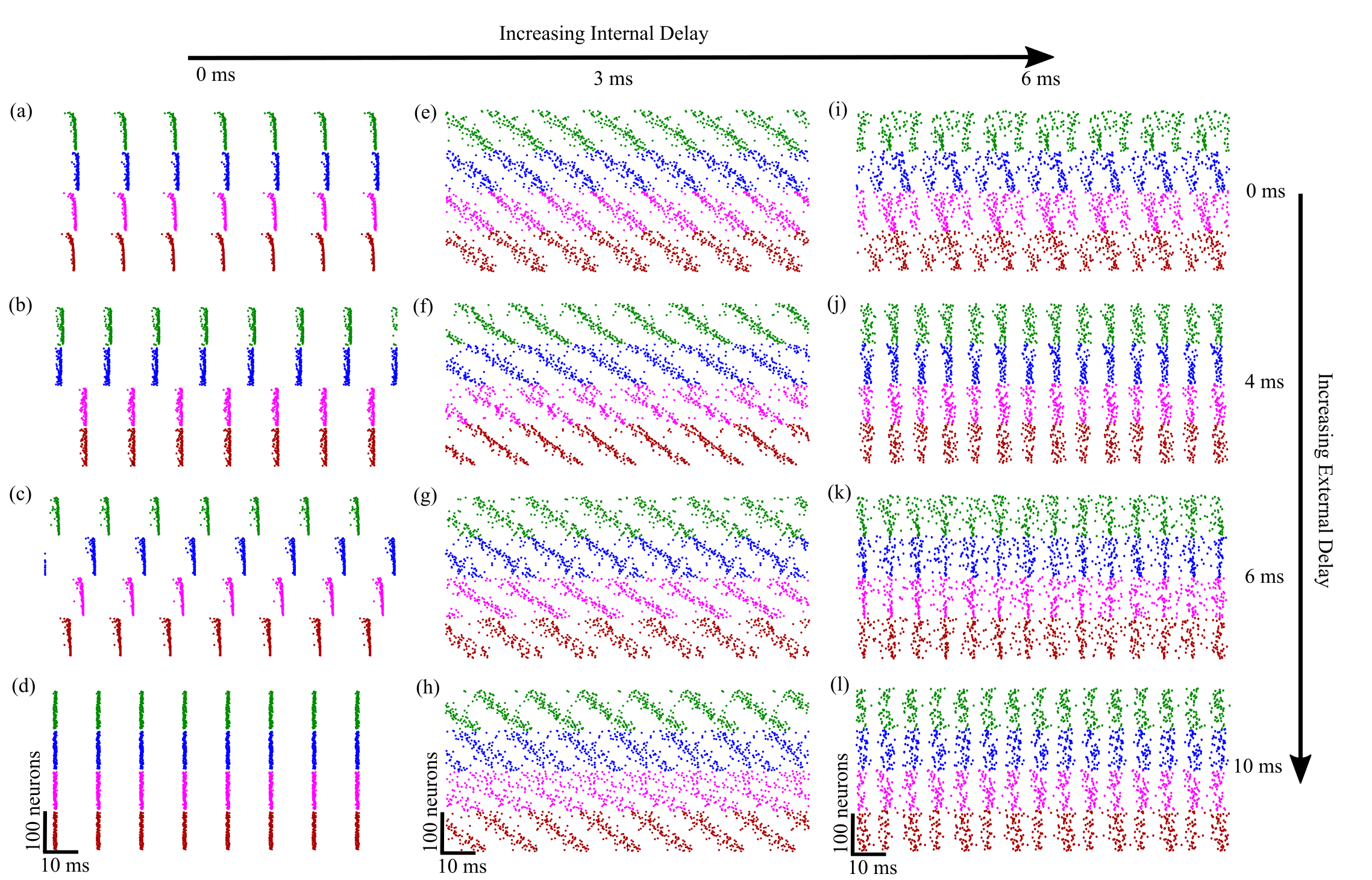}
		\caption{Raster plots of all neurons $i$ for fixed internal ($\tau_{\rm int}$) and
			external ($\tau_{\rm ext}$) time delays when synaptic plasticity is active. 
			Different colours denote each subnetwork.
			Figures (a-d), (e-h), and (i-l) display the raster plots for the internal time delays 
			equal to $\tau_{\rm int}=0$ ms, $\tau_{\rm int}=3$ ms, and $\tau_{\rm int}=6$ ms, respectively. 
			Figures (a,e,i), (b,f,j), (c,g,k), (d,h,l) display the raster plots for external 
			time delays equal to $\tau_{\rm ext}=0$ ms, $\tau_{\rm ext}=4$ ms, $\tau_{\rm ext}=6$ ms, and $\tau_{\rm ext}=10$ ms,
			respectively. For small internal delays, high synchronous patterns are observed in 
			each subnetwork, while bigger internal ones generate less synchronised patterns.}
		\label{fig2}
	\end{center}
\end{figure*}

In order to quantify and distinguish the different symmetric synchronisation
patterns, we calculate the so-called $m$-th moment of the order parameter $R^m$
($m$ is an index), that is a variation of Eq. (\ref{Rtempo}) with $m=1,2,...,S$
\cite{Sepulchre,Lucken2013,Jain} 
\begin{equation}
	R^m=\left|\frac{1}{N}\sum_{j=1}^{N}{\rm e}^{{\rm i}m\phi_j(t)}\right|.
	\label{Rm}
\end{equation} 
The summation in Eq. (\ref{Rm}) considers the phases of all neurons in the network. 
This measure allows us to quantify the number of synchronised neuronal groups
and consequently the symmetry of their phase distributions.
The calculation of the $m-th$ moment is similar to the traditional order parameter 
with the difference that considers $m=$1, 2, 3, or 4, multiplying each neuron phase. 
For the particular case where $m$=1, Eq. (17) is the same as Eq. (13). 
For the network composed of $S=4$ subnetworks, we calculate all the moments $m\in[1,4]$.
The moment $R^m$ with the highest intensity (closer to 1) provides information 
about how the subnetworks are synchronised among them. They can vary from one 
big synchronised group to two or four groups, showing a fix phase difference 
among them. In this framework, the highest moment of order parameter gives us 
the information about synchronisation and type of symmetry configuration. For
instance, if $R^1$ ($m=1$) has the highest value (close to 1), the subnetworks
have neurons forming effectively a single large network ( small phase 
difference between
subnetworks). If $R^2$ ($m$=2) is the highest value, the neurons in subnetworks 
are synchronised in $2$ groups in an anti-phase pattern (phase difference of $\pi$).
The same idea applies for $m=3$ and $m=4$, where there are 3 and 4 groups, and the 
neurons in the groups have approximately $2\pi/3$ and $\pi/2$ phase differences, respectively. 
Table \ref{table}
exhibits the standard range of parameters that we consider in our simulations.
\begin{table}[htbp]
	\caption{Descriptions of the standard parameters and range values considered in
		our simulations.} 
	\centering
	\begin{tabular}{lcc}
		\hline
		\small
		\footnotesize \bf{Descriptions} & \footnotesize \bf{Parameter} &\footnotesize
		\bf{Value} \\
		\hline
		\footnotesize Number of subnetworks & \footnotesize $S$ & \footnotesize 4 \\
		\footnotesize Neurons per subnetwork & \footnotesize $N_{\rm sub}$ &
		\footnotesize $100$ \\
		\footnotesize Internal connect. probab. &\footnotesize $p_{\rm int}$ &
		\footnotesize $1.0$ \\
		\footnotesize External connect. probab. &\footnotesize $p_{\rm ext}$ &
		\footnotesize $0.05$ \\		
		\footnotesize Internal time delay &\footnotesize $\tau_{\rm int}$ & \footnotesize $[0,6]$ ms \\
		\footnotesize External time delay &\footnotesize $\tau_{\rm ext}$ & \footnotesize $[0,12]$ ms\\
		\footnotesize Exc. synaptic conductance & \footnotesize $g_{\rm exc}$ &
		\footnotesize $[0,0.01]$ mS/cm$^2$\\
		\footnotesize Membrane capacity &\footnotesize $C$ & \footnotesize 1.0 $\mu$F/cm$^2$ \\
		\footnotesize Potassium conductance & \footnotesize$g_{\rm K}$ &  \footnotesize	36 mS/cm$^2$ \\
		\footnotesize Sodium conductance &\footnotesize $g_{\rm Na}$ & \footnotesize 120 mS/cm$^2$ \\
		\footnotesize Leak conductance &\footnotesize $g_{\rm l}$ & \footnotesize 0.3 mS/cm$^2$ \\
		\footnotesize Potassium rev. potential  &\footnotesize $V_{\rm K}$ &
		\footnotesize -77 mV \\
		\footnotesize Sodium reversal potential  &\footnotesize $V_{\rm Na}$ &
		\footnotesize 50 mV \\
		\footnotesize Leak reversal potential &\footnotesize $V_{\rm l}$ & \footnotesize -54.4 mV \\
		\footnotesize Excitatory reversal potential &\footnotesize $V_{\rm r}^{\rm+}$&
		\footnotesize 20 mV \\
		\footnotesize Constant current &\footnotesize $I_{i}$ & \footnotesize [10,11] $\mu$A/cm$^2$ \\
		\footnotesize Change rate of synap. weight & \footnotesize $G$ & \footnotesize $10^{-5}$ mS/cm$^2$. \\
		\footnotesize Time step integration & \footnotesize $\delta t$ & \footnotesize $10^{-2}$ ms \\
		\footnotesize Initial time for analyses &\footnotesize $t_{\rm ini}$ &
		\footnotesize 80 s \\
		\footnotesize Final time for analyses &\footnotesize $t_{\rm fin}$ &
		\footnotesize 100 s \\
		\footnotesize Internal time delay &\footnotesize $d_{\rm int}$ & \footnotesize	[0,1] ms \\
		\footnotesize External time delay &\footnotesize $d_{\rm ext}$ & \footnotesize	[0,12] ms \\
		\footnotesize Time step integration &\footnotesize $\delta t$ & \footnotesize	$10^{-2}$ ms \\
		\hline
	\end{tabular}
	\label{table}
\end{table}

\begin{figure*}[ht!]
	\begin{center}
		\includegraphics[scale=0.45]{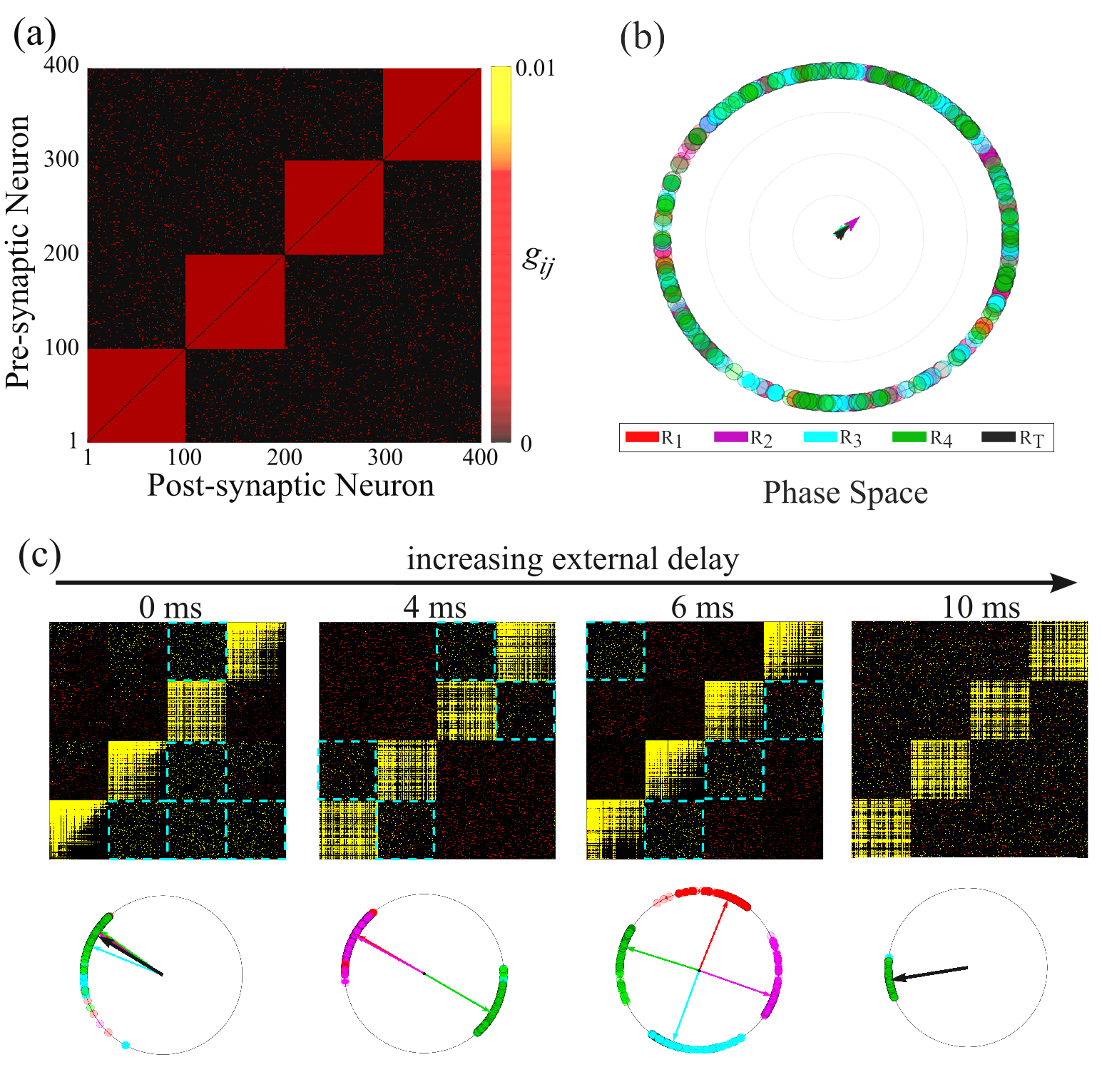}
		\caption{The panel (a) displays the initial matrix configuration for all
			simulations, where the red colour represents the initial intensity of non null synaptic connections. 
			Black indicates no synapses and yellow approximates to the maximal synaptic conductance 
			$g_{ij}^{\rm max}$=0.01 mS/cm$^2$. The panel (b) shows the initial phases of all neurons in the network. 
			The top panel in (c) exhibits the resultant matrix for fixed internal and external time delays. 
			In bottom in panel (c), the traditional Kuramoto order parameter for the entire network is 
			represented by a black arrow ($R_{\rm T}(t)$), as well as the same order parameter considering 
			each subnetwork is denoted by the red, violet,cyan, and green arrows ($R_{(S)}$).}
		\label{fig3}
	\end{center}
\end{figure*}

\section{Results and Discussions}
\label{Resuls}
In this work, we consider internal ($\tau_{\rm int}$) and external
($\tau_{\rm ext}$) delays in the subnetworks with and without the presence of
STDP. Without plasticity, for small $\tau_{\rm int}$ and varying $\tau_{\rm ext}$, 
we observe different patterns of synchronisation between the subnetworks.
However, our main goal is to investigate how these patterns of synchronisation 
affect the weights of network connections when plasticity is active. To do this, 
we consider different delay values which modify the dynamics of the network. 
We restricted the network to only excitatory neurons since the presence of 
inhibition in the current simulation generates very similar results.

Figure \ref{fig2} shows the raster plots for fixed internal and external time
delays ($\tau_{\rm int}$ and $\tau_{\rm ext}$) when the synaptic plasticity is on.
In the left side, we consider $\tau_{\rm int}=0$ ms, (a) $\tau_{\rm ext}=0$ ms, (b)
$\tau_{\rm ext}=4$ ms, (c) $\tau_{\rm ext}=6$ ms, and (d) $\tau_{\rm ext}=10$ ms. For
small internal time delays, we verify synchronised symmetric patterns. In the
center column, we consider $\tau_{\rm int}=3$ ms, (e) $\tau_{\rm ext}=0$ ms, (f)
$\tau_{\rm ext}=4$ ms, (g) $\tau_{\rm ext}=6$ ms, and (h) $\tau_{\rm ext}=10$ ms. For
these parameters, we observe no firing coherence, but the fastest neurons
(higher $I_i$) in each subnetwork start firing and subsequently the slower
neuron. In the right side, we use $\tau_{\rm int}=6$ ms, (i) $\tau_{\rm ext}=0$ ms, 
(j) $\tau_{\rm ext}=4$ ms, (k) $\tau_{\rm ext}=6$ ms, and (l) $\tau_{\rm ext}=10$ ms. 
Although some synchronisation can be noticed, it is lower than in the case for
$\tau_{\rm int}=0$ ms. For $\tau_{\rm int}=0$ ms, we identify an equal pattern for
the case with and without plasticity, as shown in Figure \ref{fig2}(a-d).

We focus on the most synchronised symmetric patterns. In Figure \ref{fig2}(a),
neuron spikes in a single group (almost complete phase synchronisation) without 
delay between the subnetworks. Highest order parameter is the one with order $m$=1, 
indicating all neurons spiking nearly synchronously. 
Figure \ref{fig2}(b) displays neurons spiking in two 
groups for an external time delay equal to $d_{\rm ext}=4$ ms. 
Highest order parameter is the one with order $m$=2. In Figure
\ref{fig2}(c), the neurons spike in four groups for an external time delay equal
to $d_{\rm ext}=6$ ms. Highest order parameter is the one with order m=4.  
For a delay close to the average period between spikes ($\approx 14$ ms) 
the network returns to a single group, as shown in Figure
\ref{fig2}(d). In this case, the neurons of all subnetworks exhibit a strong
phase synchronisation. These patterns can also be obtained for different
parameters when there is no plasticity.

Figure \ref{fig3}(a) displays the initial matrix connections $g_{ij}$ of the
subnetworks. In Figure \ref{fig3}(b), we plot the initial neuronal phases where
each colour represents a subnetwork. The coloured arrows are the initial order
parameters showing that initially the neurons are not synchronised. Figure
\ref{fig3}(c) exhibits the final coupling matrix after the plasticity actuates
by $100$ s, considering different values of the external delays and fixed 
internal delay $\tau_{\rm int}=0$ ms. For $\tau_{\rm ext}=0$ ms, we see a strong 
coherent dynamics among the subnetworks, all organised in a single group. The
resulting network presents some hierarchical organisation where the directed
connections between some subnetworks were reinforced as highlighted by the
dashed blue squares in Figure 3(c) (first column). For $\tau_{\rm ext}=4$ ms, the
network effectively forms two pairs of subnetworks with neurons belonging to 
different subnetworks having a constant phase difference around
$\pi$ radians (anti-phase synchronisation). In this case, connections between
in-phase subnetworks are potentiated and depressed for the anti-phase ones. For
$\tau_{\rm ext}=6$ ms, the network effectively presents four groups where 
neurons within each pair of groups have a phase difference around
$\pi/2$ radians and the network is set to a configuration of subnetworks being 
connected under a ring topology. Considering $\tau_{\rm ext}=10$ ms, the network 
goes effectively to one phase group formation, and the network topology shows 
no preferential connection among the subnetworks. Therefore, when the delay 
values come close to the time period of the spikes, the behaviour is similar to 
that observed when the connection delay between the subnetworks is close to zero 
and the synchronisation is improved. As the difference for one group synchronisation 
with small time delays, all connections between the 
subnetworks are potentiated. 

\begin{figure}[ht!]
	\begin{center}
		\includegraphics[scale=0.08]{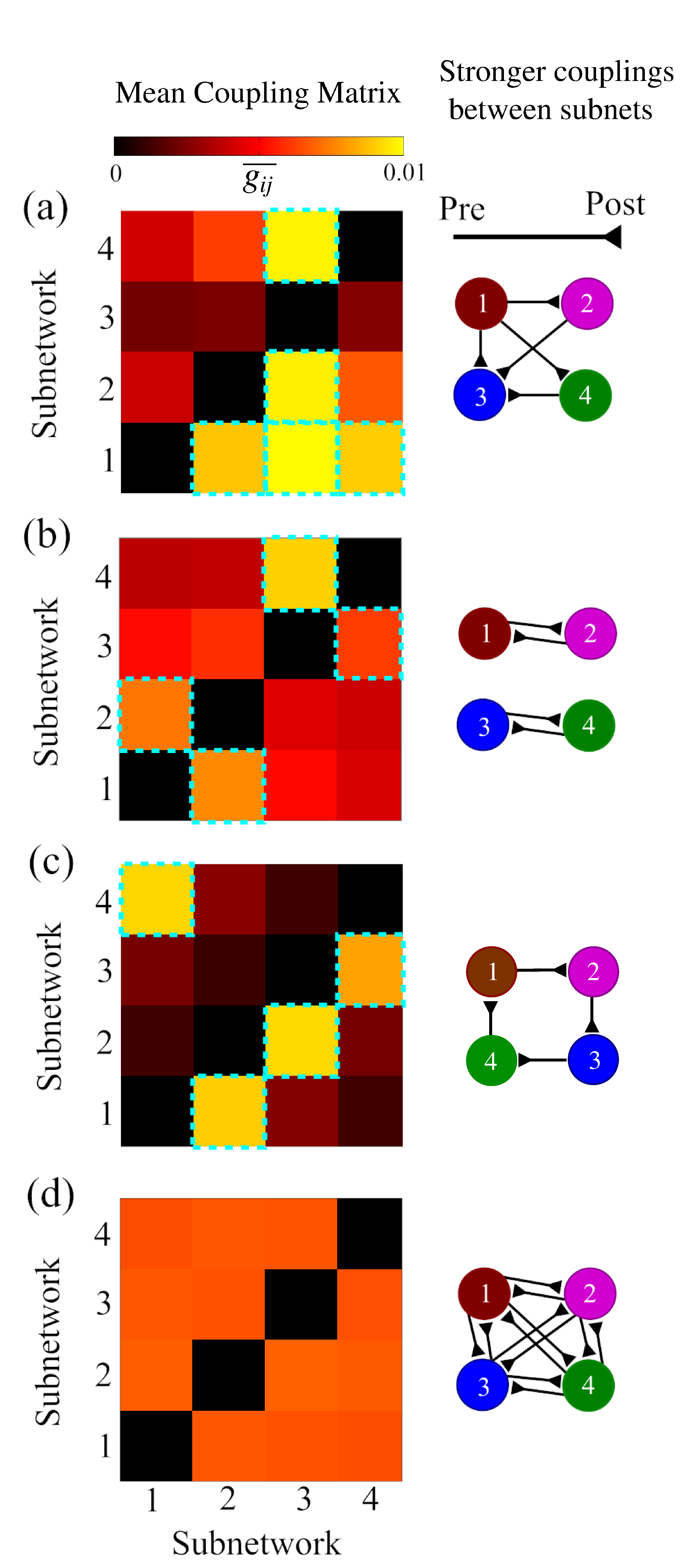}
		\caption{Resultant mean synaptic weight (left) and schematic representation of
			resultant connections between the subnetworks (right) for $\tau_{\rm int}=0$ ms.
			The edges in the graphs plotted on the right column panel represent the presence of 
			strong connections from one subnetwork to another.  
			We consider (a) $\tau_{\rm ext}=0$ ms, (b) $\tau_{\rm ext}=4$ ms, (c)
			$\tau_{\rm ext}=6$ ms and (d) $\tau_{\rm ext}=10$ ms.}
		\label{fig4}
	\end{center}
\end{figure}

In the synchronised regimes, the phase difference between groups is roughly 
constant, as seen in Figure 3(c). The traditional Kuramoto order parameter is 
not capable to detect and classify all those synchronised configurations in addition 
to a single group. For this reason, we consider the $m$-th moments of the order parameter.
The moments of the order parameter is a suitable diagnostic of symmetry between
the synchronised subnetworks.

\begin{figure*}[ht!]
	\begin{center}
		\includegraphics[scale=0.072]{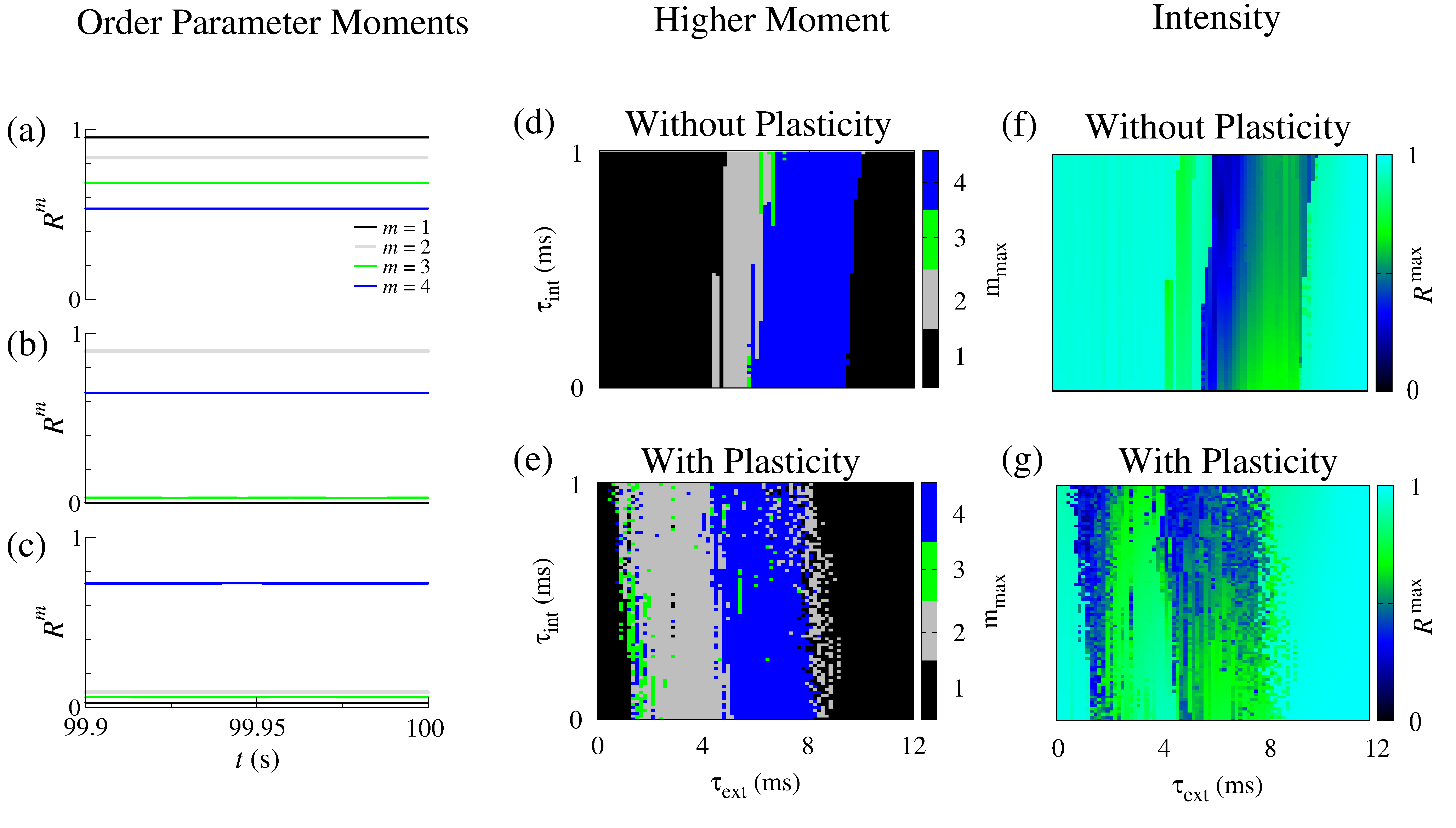}
		\caption{The panels (a) and (b) display the symmetric patterns found
			in the parameter space of $\tau_{\rm int}$ and $\tau_{\rm ext}$ without and with plasticity. 
			The panels (c) and (d) show the highest value of the order parameters moments 
			in the space parameter of $\tau_{\rm int}$ and $\tau_{\rm ext}$ for the case
			without and with synaptic plasticity.}
		\label{fig5}
	\end{center}
\end{figure*}

The left side of Figure \ref{fig4} shows the mean synaptic coupling between the
subnetworks computed with the matrices from Figure 3(c). The schematic
representations of the stronger mean connection between the subnetworks are
displayed on the right side. The strongest weight connections between the 
subnetworks can be associated to the average matrix, as highlighted by the blue
dashed squares. The black line with a triangle at the end corresponds to the 
connection direction from a presynaptic subnetwork to a postsynaptic one. In
Figure \ref{fig4}(a), the strong phase synchronisation potentiates connections
between the subnetworks in an asymmetric way. In this case, the neurons in the
subnetwork $1$ spike first, followed by subnetworks 2, 4, and 3.  STDP is
responsible for reshaping the network leading to the configurations depicted in
Figure 4(a), where the connection reinforcement follows the order of spikes. All
connections from subnetwork 1 to 2, 3, and 4 are reinforced. The same happens
with the connections from subnetwork 2 to 3 and 4, and from subnetwork 4 to 3.
Figure \ref{fig4}(b) exhibits two groups in anti-phase. The connections are
reinforced between in-phase subnetworks and weakened between the two groups. In
Figure \ref{fig4}(c), the subnetworks show a phase-lock synchronisation with
average phase difference of $\pi$ radians. These patterns lead to a ring network
configuration. The connections are potentiated in a cyclic way. The subnetworks
are in-phase synchronisation in Figure \ref{fig4}(d) in a more coherent state than
in Figure \ref{fig4}(a). This strong synchronisation promotes an all-to-all
connection organisation among subnetworks with an average reinforcement of the
connections.

To better understand how plasticity promotes synchronised patterns, in Figure \ref{fig5}, 
we calculate the $m$-th moments for $\tau_{\rm int}=0$ ms, (a) $\tau_{\rm ext}=0$ ms, 
(b) $\tau_{\rm ext}=4$ ms and (c) $\tau_{\rm ext}=6$ ms. To calculate the $R^m$, 
we consider the phase temporal evolution of all neurons of the network in Equation \ref{Rm}, 
independent of their respective subnetwork, for each moment $m$ = 1, 2, 3, and 4.
We observe a higher 1-st
moment in Figure \ref{fig5}(a), which corresponds to a one-group 
phase synchronisation between the subnetworks. In this case, we verify that
other moments are relatively lower than the first one. Figure \ref{fig5}(b)
displays a higher 2-nd moment due to an anti-phase synchronisation in two major
groups. In Figure \ref{fig5}(c), we identify a higher 4-th moment, which is
associated with a phase-lock synchronisation between the subnetworks, as shown
in Figure \ref{fig3}(c). The higher moment of the order parameter indicates the
symmetry of the synchronised patterns. It is worth to mention, that as can be seen 
in Figures \ref{fig3}(a-c), the moments present a constant value overtime.
Figures \ref{fig5} (d) and (e) exhibit the 
symmetric synchronised patterns in the parameter space $\tau_{\rm ext}\times\tau_{\rm int}$ 
without and with synaptic plasticity, respectively. In the case without plasticity 
and small external time delay, we find one group symmetry with phase synchronisation. 
Increasing the external delays, we identify predominantly 2 and 4 groups symmetry, 
respectively. For larger external delays (around 10 ms), the network returns to a 
one synchronised group. These cases are found for small internal time delays (less than 1 ms). 
For higher internal time delays (more than 1 ms), we observe less synchronised patterns.
On the other hand, synaptic plasticity shifts to the left in $\tau_{\rm ext}$, 
in which the regions where one, two, and four group symmetries are found. 
Moreover, the area in parameter space covering 
the existence of order 2 and 4 phase patterns is enlarged by plasticity. 
Then, not only plasticity allows for more complex patterns to emerge for smaller $\tau_{\rm ext}$, 
but also the livelihood of its appearance for a large range of delays. 
Thus, plasticity promotes complexity as measured by the emergence of symmetric 
synchronous patterns. Figures \ref{fig5}(f) and \ref{fig5}(g) show the higher 
values of the order parameters moments for the case without and with synaptic 
plasticity, respectively. We verify that the highest values of $R^m$ correspond 
to one group symmetry.

In summary, spiking synchronous patterns are strongly related to the network 
connection between subnetworks in a plastic neuronal network with time delay.
For small external delays, one group configuration exhibits subnetworks spiking 
in a specific order, and this order promotes synaptic potentiation from the subnetworks 
spiking first to the subsequent ones. Increasing external delay, two group patterns 
generate potentiation in in-phase subnetworks, while a synaptic depression 
among the ones from distinct groups. For major delays, four groups (a phase-lock 
synchronisation between subnetworks) show a synchronised state, in which all subnetworks 
spiking almost at the same time and without any preferential order. 
This dynamical behaviour promotes a final global network configuration with an 
average increase in all synaptic strengths. In all cases, the potentiation between 
the neuronal areas can also depend on the internal time delay. The less recurrent 
pattern found in our simulations is the 3 groups organisation, which can be 
associated with a transient behaviour.

%%%%%%%%%%%%%%%%%%%%%%%%%%%%%%%%%%%%%%%%%%%%%%%
%%%%%%%%%%%%%%%%%%%%%%%%%%%%%%%%%%%%%%%%%%%%%%%

\section{Conclusions}
\label{sec_conclusions}

In this work, we consider a network of subnetworks to study the effects of 
internal and external transmission delays on the generation of symmetric
dynamics patterns, as well as the potentiation and depression on the synaptic
weights due to the presence of plasticity. To do that, we consider
Hodgkin-Huxley neurons coupled by means of excitatory chemical synapses and a
time dependent plastic rule. To achieve synchronised patterns, the internal
delayed transmission of neuron communication in the subnetworks assumes small 
values while the external
one assumes higher values. When the internal transmission delay of neurons in the
subnetworks assumes higher values, nonsynchronised patterns are observed.

Without plasticity and depending on the delay transmission between the
subnetworks, we verify that synchronisation among subnetworks can be observed 
in different patterns. These regimes can be detected by means of the $m$-th 
moment of the order parameter, which provides the information about how the
dynamics of neurons in the subnetworks are correlated in the phase space of 
spiking times. Due to the plasticity effect, we verify that the final connectivity 
reflects the symmetric synchronous patterns on the emergence of the strongest 
connections between the subnetworks. Thus, subnetworks that are strongly connected, 
are also strongly synchronous, and the same symmetric patterns observed in 
terms of synchronisation are also found in the final connecting topology.
We show that the synaptic transmission delays play an important role in the
generation of symmetric synchronised patterns. In addition, we also show that
the phase, anti-phase, and symmetric phase-lock synchronised firing
patterns influence the synaptic changes of the weight connections among the
subnetworks. We observe the relationship of each symmetric synchronised firing pattern 
with the induced potentiation between the subnetworks. As a consequence, 
our results suggest that firing patterns can induce different topologies 
in addition to that topology induced firing patterns. We show that it is 
possible to identify spiking correlations and delayed excitatory connectivities 
between different neuronal groups.

A recent work \cite{Giannari2020} has proposed the construction of modular, 
scalable and adaptable neuronal networks with neurons possessing different 
dynamic properties and distinct firing patterns. To achieve that the authors 
considered neurons with different external input. In this work, we also consider 
heterogeneity in the external parameter. Our goal however was focused not 
on the design but on the relation between topology and synchronisation patterns. 
Whenever networks with STDP are designed with neurons that can have external 
currents controllable and accessible, our work shows that topology and 
synchronisation are strongly related.

Topology and behaviour is extensively studied in the literature. The novelty 
in our results was to study the emergency of different synchronous regimes, 
phase, anti-phase and shift phase. The emergency of these distinct regimes 
has been found in the brain. Experimental evidence shows that phase synchronisation 
is able to support memory processes and changes in the connection strengths \cite{fell2011,Clouter2017}. 
In particular, results of Jutras and Buffalo suggest that phase synchronisation 
can lead to potentiation of the synaptic connections \cite{Jutras2010}. 
Our results can be linked to these experimental works, providing a possible 
explanation about how these synchronous firing patterns leads to a topology. 
Anti-phase oscillations are observed during rest state in humans \cite{Fox2007} 
and in anaesthetised monkeys \cite{Vicent2007}. In addition, shift-phase 
synchronisation was observed in primary visual cortex \cite{Vinck2010} and could 
play a function to stimulus selection \cite{Tiesinga2010}. Our work shows that 
all these synchronous phenomena induce topology and emerge from it.

All these paving the way for us to conclude that plasticity - described by a pairwise 
function that regulates synapse strength by the time intervals between two spiking neurons - 
in fact promotes the creation of evolved network structures whose subnetworks of intra 
connected neurons and their inter connections is strongly reflected in the global 
synchronisation patterns measured by the phase dynamics of the neurons. Thus, 
the plastic neural network has a strong match between phase activity and graph structure.     

\section*{Conflict of Interest Statement}

The authors declare that there is no conflict of interest.

\section*{Acknowledgements}

The authors acknowledge the financial support from S\~ao Paulo Research
Foundation (FAPESP, Brazil) (Grants Nos. 2016/23398-8, 2017/13502-5, 2018/03211-6, 
2020/04624-2, 2022/05153-9, 2022/13761-9), 
National Council for
Scientific and Technological Development (CNPq), Funda\c c\~ao Arauc\'aria and 
Coordena\c c\~ao de Aperfei\c coa\-mento de Pessoal de N\'ivel Superior -
Brasil (CAPES).

%%%%%%%%%% References %%%%%%%%%%%%%%

\bibliographystyle{elsarticle-harv}

\begin{thebibliography}{00}
	\bibitem{petkoski2019}
	Petkoski S, Jirsa VK. Transmission time delays organize the brain network synchronization. 
	Philos. Trans. Royal Soc. A 2019; 377: 20180132.	
	\bibitem{sreenivasan2019}
	Sreenivasan KK, D'Esposito M. The what, where and how of delay activity. 
	Nat. Rev. Neurosci. 2019; 20: 466.
	\bibitem{Asl2018}
	Asl MM, Valizadeh A, Tass PA Delay-induced multistability and loop formation in 
	neuronal networks with spike-timing-dependent plasticity. Sci Rep, 2018; 8: 12068.	
	\bibitem{borges2022} Borges FS, Moreira JVS, Takarabe LM, Lytton WW, Dura-Bernal S. 
	Large-scale biophysically detailed model of somatosensory thalamocortical circuits in NetPyNE. 
	Front. Neuroinform. 2022; 16: 884245.
	\bibitem{Stuart1997} Stuart G, Schiller J, and Sakmann B. Action potential initiation 
	and propagation in rat neocortical pyramidal neurons. J. Physiol. 1997; 505: 617-632.
	\bibitem{Kosuke2022} Itoh K, Konoike N, Nejime M, Iwaoki H, Igahashi H, Hirata S, 
	Nakamura K. Cerebral cortical processing time is elongated in human brain evolution. 
	Sci. Rep. 2022; 12: 1103.
	\bibitem{lameu18}
	Lameu EL, Macau EEN, Borges FS, Iarosz KC, Caldas IL, Borges RR, Protachevicz PR, 
	Viana RL, Batista AM. Alterations in brain connectivity due to plasticity and synaptic 
	delay. Eur. Phys. J. 2018; 227:673-682.
	\bibitem{Mugnaine2018} 
	Mugnaine M, Reis AS, Borges FS, Borges RR, Ferrari FAS, Iarosz KC, Caldas IL, 
	Lameu EL, Viana RL, Szezech Jr JD, Kurths J, Batista AM.
	Delayed feedback control of phase synchronisation in a neuronal network model. 
	Eur. Phys. J. Spec. Top. 2018; 227: 1151-1160.
	\bibitem{Hansen2022}
	Hansen M, Protachevicz PR, Iarosz KC, Caldas IL, 
	Batista AM, Macau EEN. The effect of time delay for 
	synchronization suppression in neuronal networks. 
	Chaos Solit. Fractals 2022; 164: 112690.
	\bibitem{lubenov2008}
	Lubenov EV, Siapas AG. Decoupling through synchrony in neuronal circuits with 
	propagation delays. Neuron 2008;58:118-131.
	\bibitem{protachevicz2020a}
	Protachevicz PR, Borges FS, Iarosz KC, Baptista MS, Lameu EL, Hansen M, Caldas IL, 
	Szezech Jr. JD, Batista AM, Kurths J. Influence of delayed conductance on neuronal 
	synchronization. Front. Physiol. 2020; 11: 1-9.
	\bibitem{Power2011}
	Power JD, Cohen AL, Nelson SM, Wig GS, Barnes KA, Church JA, Vogel AC, Laumann TO, 
	Miezin FM, Schlaggar BL, Petersen S. Functional network organization of the human brain. 
	Neuron. 2011; 72(4): 665-78.
	\bibitem{SpornsBetzel2015}
	Sporns O, Betzel RF. Modular Brain Networks. Annu. Rev. Psychol. 2016; 67: 613-40.
	\bibitem{Lin2013}
	Lin F-H, Witzel T, Raij T, Ahveninen J, Tsai KW-K, Chu Y-H, Chang W-T
	Nummenmaa A,  Polimeni JR, Kuo W-J,  Hsieh J-C, Rosen BR,  Belliveau JW. fMRI 
	hemodynamics accurately reflects neuronal timing in the human brain measured by MEG. 
	Neuroimage. 2013; 78: 372-384.
	\bibitem{Guo2022}
	Guo B, Zhou F, Li M, Gore JC. Latency structure of BOLD signals within white matter 
	in resting-state fMRI. Magnetic Resonance Imaging. 2022; 89: 58-69.
	\bibitem{Sun2018}
	Sun X,  Perc M, Kurths J, Lu Q. Fast regular firings induced by intra- and inter-time 
	delays in two clustered neuronal networks. Chaos 2018; 28: 106310.
	\bibitem{Know2009}
	Know J, Choe Y. Facilitating neural dynamics for delay compensation: a road 
	to predictive neural dynamics? Neural Netw. 2009; 22: 267-276.
	\bibitem{ramirez2016}
	Ramirez A, Arbuckle MR. 
	Synaptic plasticity: the role of learning and unlearning in addiction and 
	beyond. Biol. Psychiatry 2016; 80: e73.
	\bibitem{abraham2019} 
	Abraham WC, Jones OD, Glanzman DL. Is plasticity of synapses
	the mechanism of long-term memory storage? NPJ Sci. Learn. 2019; 4: 9.
	\bibitem{fell2011} 
	Fell J, Axamacher N. The role of phase synchronization in memory processes. 
	Nat. Rev. Neurosci. 2011; 12:105.
	\bibitem{Kim2018a}
	Kim S-Y, Lim W. Stochastic spike synchronization in a small-world 
	neural network with spike-timing-dependent plasticity. Neural Netw. 
	2018; 97: 92-106.
	\bibitem{Kim2018b}
	Kim S-Y, Lim W. Effect of inhibitory spike-timing-dependent plasticity on 
	fast sparsely synchronized rhythms in a small-world neuronal network.
	Neural Netw. 2018; 106: 50-66.
	\bibitem{Kim2021}
	Kim S-Y, Lim W. Effect of diverse recoding of granule cells on optokinetic 
	response in a cerebellar ring network with synaptic plasticity. 
	Neural Netw. 2021; 134: 173-204.
	\bibitem{Soltoggio2012}
	Soltoggio A, Stanley KO. From modulated Hebbian plasticity to simple behavior 
	learning through noise and weight saturation. Neural Netw. 
	2012; 34: 28-41.
	\bibitem{Aoki2015}
	Aoki T. Self-organization of a recurrent network under ongoing 
	synaptic plasticity. Neural Netw., 2015; 62: 11-19. 
	\bibitem{tognoli2009} 
	Tognoli E, Kelso JAS. Brain coordination dynamics: true and false
	faces of phase synchrony and metastability. Prog. Neurobiol. 2009; 87: 31.	
	\bibitem{thatcher2012} 
	Thatcher RW. Coherence, phase differences, phase shift, and phase lock
	in EEG/ERP analyses. Dev. Neuropsychol. 2012; 37: 476-496.
	\bibitem{carlos2020} 
	Carlos F-LP, Ubirakitan M-M, Rodrigues MCA, Aguilar-Domingo M, Herrera-Gutiérrez E, 
	Gomez-Amor J, Coppelli M, Carreli PV, Matias FS. 
	Anticipated synchronization in human EEG data: Unidirectional 
	causality with negative phase lag. Phys. Rev. E 2020; 102: 032216.
	\bibitem{Protachevicz2021}
	Protachevicz PR, Hansen M, Iarosz KC, Caldas IL, Batista AM, Kurths J. Emergence of 
	neuronal synchronization in coupled areas. Front. Comput. Neurosci. 2021; 15: 1-12.
	\bibitem{klimesch2008} 
	Klimesch W, Freunberger R, Sauseng P, Gruber W. A short review of
	slow phase synchronization and memory: evidence for control processes in
	different memory systems? Brain Res. 2008; 1235: 31-44.
	\bibitem{knoublauch2003} 
	Knoublauch A, Sommer FT. Spike-timing-dependent synaptic plasticity
	can form ``zero lag links'' for cortical oscillations. Neurocomputing 2004; 52-54:
	301-306.
	
	\bibitem{li2011} 
	Li D, Zhou C. Organization of anti-phase synchronization pattern in
	neural networks: what are the key factors? Front. Syst. Neurosci. 2011; 5: 1-14. 
	\bibitem{bodner1997} 
	Bodner M, Zhou YD, Shaw GL, Fuster JM. Symmetric temporal
	patterns in cortical spike trains during performance of a short-term memory
	task. Neurol. Res. 1997; 19: 509-514. 
	\bibitem{Manor1991} 
	Manor Y, Koch C, Segev I. Effect of geometrical irregularities on propagation delay in 
	axonal trees. Biophys. J. 1991, 60: 1424-1437.
	\bibitem{Boudkkazi2007}
	Boudkkazi S, Carlier E, Ankri N, Caillard O, Giraud P, Fronzaroli-Molinieres L, Debanne D. 
	Release-dependent variations in synaptic latency: a putative code for short-and long-term 
	synaptic dynamics. Neuron 2007, 56: 1048-1060. 
	\bibitem{Wang2005}
	Wang HX, Gerkin RC, Nauen, DW, Bi GQ. Coactivation and timing-dependent integration 
	of synaptic potentiation and depression. Nat. Neurosci. 2005, 8: 187-193. 
	\bibitem{Knoblauch2003}
	Knoblauch A, Sommer FT. Synaptic plasticity, conduction delays, and inter-areal phase 
	relations of spike activity in a model of reciprocally connected areas. Neurocomputing 2003, 52: 301-306. 
	\bibitem{AgmonSnir1993}
	Agmon-Snir H, Segev I. Signal delay and input synchronization in passive dendritic 
	structures. J. Neurophysiol. 1993, 70: 2066-2085. 
	\bibitem{Schierwagen2001}
	Schierwagen A, Claus C. Dendritic morphology and signal delay in superior colliculus 
	neurons. Neurocomputing 2001. 38: 343-350. 
	\bibitem{Swadlow1987}
	Swadlow HA, Weyand TG. Corticogeniculate neurons, corticotectal neurons, and suspected 
	interneurons in visual cortex of awake rabbits: receptive-field properties, axonal
	properties, and effects of eeg arousal. J. Neurophysiol. 1987, 57: 977-1001. 
	\bibitem{hodgkin52}
	Hodgkin AL, Huxley AF. A quantitative description of membrane
	current and its application to conduction and excitation in nerve. Physiol. J. 1952; 11: 500.
	\bibitem{Luccioli2006}	
	Luccioli S, Kreuz T, Torcini A. Dynamical response of the Hodgkin-Huxley model in the 
	high-input regime. Phys. Rev. E 2006, 73: 041902.
	\bibitem{Pospischil2008}
	Pospischil M, Toledo-Rodriguez M., Monier C, Piwkowska Z, Bal T, Frégnac Y, Markram H, 
	Destexhe A. Minimal Hodgkin-Huxley type models for differents classes of cortical and 
	thalamic neurons. Biol. Cybern. 2008; 99: 427-411.
	\bibitem{Giannari2020}
	Giannari A G, Astolfi A. Model design for networks of heterogeneous Hodgkin-Huxley neurons. 
	Neurocomputing. 2020; 496: 147-157. 
	\bibitem{Shi2016}
	Shi Q, Han F, Wang Z, Li C. Rhythmic oscillations of excitatory bursting Hodgkin-Huxley 
	neuronal network with synaptic learning. Comput. Intell. Neurosci. 2016; 6023547: 1-9.
	\bibitem{Popovych2013}	
	Popovych OV, Yanchuk S, Tass PA. Self-organized noise resistance of oscillatory 
	neural networks with spike timing-dependent plasticity. Sci. Rep. 2013, 3: 2926.
	\bibitem{Borges2015}
	Borges RR, Iarosz KC, Batista KC, Caldas IC, Borges FS, Lameu EL. Sincronização de 
	disparos em redes neuronais com plasticidade sináptica. Rev. Bras. Ensino Fis. 2015, 37(2): 2310.
	\bibitem{Borges2016}
	Borges RR, Borges FS, Lameu EL, Batista AM, Iarosz KC, Caldas IL, Viana RL, Sanjuán MAF. 
	Effect of the spike timing-dependent plasticity on the synchronization in a random 
	Hodgkin-Huxley neuronal network. Commun. Nonlinear. Sci. Numer. Simulat. 2016; 34: 12-22.
	\bibitem{Lameu2018}
	Lameu EL, Macau EEN, Borges FS, Iarosz KC, Caldas IL, Borges RR, Protachevicz PR, 
	Viana RL, Batista AM. Alterations in brain connectivity due to plasticity and 
	synaptic delay. Eur. Phys. J. Special Topics 2018, 227: 673-682.
	
	\bibitem{Rothman2014}
	Rothman JS, Silver RA.
	Data-driven modeling of synaptic transmission and integration. 
	Prog. Mol. Biol. Transl. Sci. 2014; 123: 305-350.
	\bibitem{borgesNN2017}
	Borges FS, Protachevicz PR, Lameu EL, Bonetti RC, Iarosz KC, Caldas IL, Baptista MS, 
	Batista AM. Synchronised firing
	patterns in a random network of adaptive exponential integrate-and-fire neuron
	model. Neural Netw. 2017; 90: 1-7.
	\bibitem{Asl2017}
	Asl MM, Valizadeh A, Tass PA. Dendritic and axonal propagation
	delays determine emergent structures of neuronal networks with plastic synapses.
	Sci. Rep. 2017; 7: 39682.
	\bibitem{Markram1997}
	Markram H,  L\"ubke J, Frotscher M, Roth A, Sakmann B.
	Physiology and anatomy of synaptic connections between thick tufted pyramidal
	neurones in the developing rat neocortex. Physiol. J. 1997; 500.2: 409-440.	
	\bibitem{Markram1997b}
	Markram H, L\"ubke J, Frotscher M, Sakmann B. Regulation of
	synaptic efficacy by coincidence of postsynaptic APs and EPSPs. Science, 1997; 275:
	5297.
	\bibitem{borges17}
	Borges RR, Borges FS, Lameu EL, Batista AM, Iarosz KC,
	Caldas IL, Antonopoulos CG, Baptista MS. Spike
	timing-dependent plasticity induces non-trivial topology in the brain. 
	Neural Netw. 2017; 88: 58.
	\bibitem{bi98}
	Bi GQ, Poo MM. Synaptic modifications in cultured hippocampal
	neurons: dependence on spike timing, synaptic strength, and postsynaptic cell
	type. J. Neurosci. Res. 1998; 18: 10464.
	\bibitem{Markram2012}
	Markram H, Gerstner W, Sj\"ostr\"om PJ. Spike-timing-dependent
	plasticity: a comprehensive overview. Front. Synaptic Neurosci. 2012; 4: 2.
	\bibitem{Caporale}
	Caporale N, Dan Y. Spike timing-dependent plasticity: a hebbian
	learning rule. Annu. Rev. Neurosci. 2008; 31: 25-46.
	\bibitem{Popovych}
	Popovych OV, Yanchuk S, Tass PA. Self-organized noise resistance
	of oscillatory neural networks with spike timing-dependent plasticity.
	Sci. Rep. 2013; 3: 2926.
	\bibitem{Protachevicz2020b}
	Protachevicz PR, Iarosz KC, Caldas IL, Antonopoulos CG, Batista AM, Kurths J. 
	Influence of autapses on synchronization in neural
	networks with chemical synapses. Front. Syst. Neurosci. 2020; 14: 604563.
	\bibitem{Schmidt}
	Schmidt M, Bakker R, Hilgetag CC, Diesmann M, van Albada SJ. 
	Multi-scale account of the network structure of macaque visual cortex. 
	Brain Struct. Funct. 2018; 223: 1409-1435.
	\bibitem{Johnson}
	Johnson RR, Burkhalter A. Microcircuitry of forward and feedback
	connections within rat visual cortex. J. Comp. Neurol. 1996; 368: 383-398.
	\bibitem{kuramoto84}
	Kuramoto Y. Chemical Oscillations, Waves, and Turbulence 1984
	(Springer-Verlag, Berlin).
	\bibitem{Sepulchre}
	Sepulchre R, Paley DA, Leonard NE. Stabilization of planar
	collective motion: All-to-all communication. IEEE Trans. Automat. Contr. 2007; 52: 5.
	\bibitem{Lucken2013}
	L\"ucken L, Yanchuk S, Popovych OV, Tass PA. 
	Desynchronization boost by non-uniform coordinated reset stimulation in 
	ensembles of pulse-coupled neurons. Front. Comput. Neurosci. 2013; 7: 63.
	\bibitem{Jain}
	Jain A, Ghose D. Collective circular motion in synchronized and
	balanced formations with second-order rotational dynamics. 
	Commun. Nonlinear Sci. Numer. Simul. 2018; 54: 156-173.
	\bibitem{Clouter2017}	
	Clouter A, Shapiro KL, Hanslmayr S. The phase synchronization is the glue 
	that binds human associative memory. Curr. Biol. 2017; 27: 3143-3148.		
	\bibitem{Jutras2010}
	Jutras MJ, Buffalo EA. Synchronous neural activity and memory formation. 
	Curr. Opin. Neurobiol. 2010; 20(2): 150-155.	
	\bibitem{Fox2007} 
	Fox MD, Raichle ME. Spontaneous fluctuations in brain
	activity observed with functional magnetic resonance imaging. Nat. Rev. Neurosci. 2007; 8: 700.	
	\bibitem{Vicent2007}
	Vincent JL, Patel GH, Fox MD, Snyder AZ, Baker JT, Van Essen DC, Zempel JM,
	Snyder LH, Corbetta M, Raichle ME. Intrinsic functional architecture in the 
	anaesthetized monkey brain. Nature 2007; 447: 83-86.
	\bibitem{Vinck2010}
	Vinck M, Lima B, Womelsdorf T, Oostenveld R, Singer W, Neuenschwander S, Fries P. 
	Gamma-phase shifting in awake monkey visual cortex. J. Neurosci. 2010; 30: 1250-1257. 
	\bibitem{Tiesinga2010}
	Tiesinga PH, Sejnowski TJ. Mechanisms for phase shifting in cortical networks 
	and their role in communication though coherence. Front. Hum. Neurosci. 2010; 4.
\end{thebibliography}
\end{document}